\documentclass[aps,prl,twocolumn,groupedaddress]{revtex4}

\usepackage{graphicx}
\usepackage{dcolumn}
\usepackage{bm}

\begin{document}

\title{Quantum Conductance of Achiral Graphene Ribbons and Carbon Nanotubes}

\author{Lyuba Malysheva and Alexander Onipko}
\thanks{This paper is published in Zh. Eksp. Teor. Fiz. {\bf 135}, 139 (2009). There is no difference in the presentation of the results between the present arX version and published paper. However, there are some small but nevertheless important textual changes in this version, where a few misprints are eliminated. Also, Eq. (7) reads as presented here, and not as in the journal version. We regret for this inconvenience caused by the editorial process which, in our opinion, could have better feadback with the authors and better timing.}
\affiliation{Bogolyubov Institute for Theoretical Physics, 03680 Kyiv, Ukraine}

\date{\today}

\begin{abstract}
Explicit expressions of the band spectrum near the neutrality point are derived for armchair and zigzag graphene ribbons and carbon tubes. Several spectral features, which were previously observed only in numerical calculations, are given an adequate analytic description in terms of elementary functions. The obtained dispersion relations are used for a comparison of conductance ladders of graphene-based wires; these relations are also beneficial for many other applications. 
\end{abstract}

\pacs{73.22.-f }

\maketitle

\section{Introduction}
For electrons and holes, graphene ribbons and carbon tubes are one-dimensional wires made of one-atom-thick material. In comparison with the two-dimensional electron gas counterparts in semiconductor heterostructures, the transport of charge carriers in graphene \cite{Geim} (in particular, quantum conductance \cite{Peres1}) demonstrates a number of unusual properties. This paper gives a precise analytic description of the conductance of four basic graphene wires and specific features of each member of the wire family represented in Fig.~1, achiral graphene ribbons and carbon tubes.  Formally, this problem can be considered already "solved" by finding two equations that describe the  spectrum of a graphene sheet with two armchair- and  two zigzag-shaped edges \cite{Lyuba1} (in the center of Fig.~1). But spectrum peculiarities near the Fermi energy \cite{Peres1,Lyuba1,Fujita,Mint,Brey,Ch,Lyuba2} are far from being obvious from the general equations. Here, we show that (i) the band spectrum of a metallic armchair (zigzag) ribbon (tube) is not the same as for zigzag (armchair) ribbon (tube); this difference is given an accurate quantitative description; (ii) in moving away from the Fermi energy, the bottoms (tops) of conduction  (valence) bands in zigzag (armchair) ribbons (tubes) are shifted towards larger wave  vectors; and (iii) there exist three types of spectra (conductance ladders) with equal, irregular, and alternating band spacing (ladder step width). By expressing each of these features in elementary  functions, the understanding achieved in previous studies is considerably improved.

\begin{figure}
\includegraphics[width=0.45\textwidth]{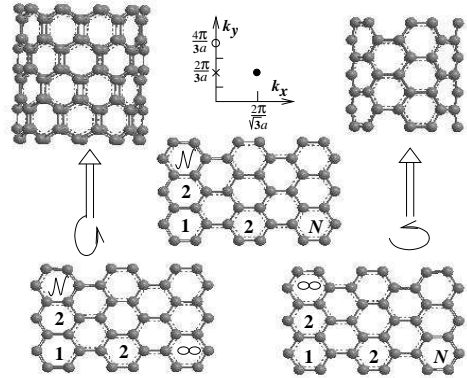}
\caption{Parent graphene $N$$\times$$\cal N$ lattice (center) and its daughter structures, from lower-left clockwise: armchair ribbon ($N=\infty$); $({\cal N},0)$ tube ($N=\infty$, 1-st = ${\cal N}+1$-th polyparaphenylene chain); $(N,N)$ tube (${\cal N}=\infty$, 1-st = $N+1$-th polyacene chain); and zigzag ribbon (${\cal N}=\infty$). In all cases, same k-coordinates are used. The cross at $(0,\frac{2\pi}{3a})$ and the circle at $(0,\frac{4\pi}{3a})$ indicate zero-energy points: {\small $\times$} for armchair metallic GR, {\small $\times$} and {\large $\circ$} for zigzag metallic CT; the filled circle at $(\frac{2\pi}{\sqrt{3}a},\frac{2\pi}{3a})$ indicates the second special point for the armchair CT. For the zigzag GR, zero-energy point cannot be shown on real $k_xk_y$ plane.}
\end{figure}

\section{Band Structure} 
Figure 1 illustrates the parent honeycomb $N$$\times$$\cal N$ lattice and its daughter wire-like structures, armchair and zigzag graphene ribbons and carbon tubes, henceforth abbreviated as GR and CT, respectively. The lattice label indicates that in the armchair direction, graphene contains $N$ hexagons in polyparaphenylene-like chains, whereas in the zigzag direction, it has ${\cal N}$ hexagons forming polyacene-like chains. Hydrogen atoms along edges are not shown and not taken into account in the nearest-neighbor tight-binding Hamiltonian \cite{Wallace,Saito}. 

The $\pi$-electron spectrum of this model is given by \cite{Lyuba1}
\begin{equation} \label{c2}
E^\pm= \pm \sqrt{1+4\cos^2\frac{ak_y}{2}\pm4\left|\cos\frac{ak_y}{2}
\cos\frac{\sqrt{3}ak_x}{2}\right|},
\end{equation}
where the hopping integral is used as an energy unit, $a$ is the minimal translation distance of the honeycomb lattice, and one of the two dimensions, $N$ or $\cal N$, is supposed to be infinite, implying the continuity of $k_x$ (for armchair GR and zigzag CT) or $k_y$ (for zigzag GR and armchair CT). The complementary discrete quantum numbers are respectively determined by open ends and periodic boundary conditions for ribbons and tubes. Thus, for graphene ribbons, the $k$ space is $(0{\rm -}\pi,0{\rm -}\pi)$; for armchair and zigzag carbon tubes, the required extensions of this space are respectively $(0­{\rm -}2\pi,0{\rm -}\pi)$ and $(0{\rm -}\pi,0{\rm -}2\pi)$. 

The spectra of armchair GR (aGR), armchair CT (aCT), and zigzag CT (zCT) are completely determined by Eq.~(\ref{c2}) and by the boundary conditions dictating values of the discrete quantum number. An additional equation, Eq.~(\ref{c8}), comes into play in the discussion of zigzag graphene ribbons.

We consider related GR and CT pairs separately. The focus is on the energies not far away from the point of neutrality, which coincides with zero energy, the Fermi energy. In this energy region, the spectrum is described by the minus branch of Eq.~(\ref{c2}) (the minus sign in the radicand). Because of the spectrum symmetry, we refer only to the conduction one-dimensional bands, that is, to the $E^-$ branch with the plus sign in front of the root. The valence bands, having the same transverse quantum numbers, are just the mirror reflection of the conduction bands in the $E=0$ plane. The analysis is performed for large $N$ and ${\cal N}$. This simplification can easily be avoided, but even for $N,{\cal N}>10$, it is sufficiently good for reasonable estimates.

{\bf Armchair ribbons and zigzag tubes}.
For both types of structures, aGR and zCT, $k_x$ is a continuous variable, $0\le\sqrt{3}ak_x\le\pi$. 
The wave-vector transverse component takes discrete values $k_y=\xi_j/a$, $\xi_j=\pi j/({\cal N}$+1), $j=1,2,...,{\cal N}$ and $\xi_j=2\pi j/{\cal N}$, $j=0,1,...,{\cal N}$$-$1 for ribbons and tubes, respectively.

It is known [and also follows from Eq.~(\ref{c2})] that an aGR (zCT) is metallic if $j^*=2({\cal N}+1)/3$ ($j^*={\cal N}/3$) is an integer. In this case, $E^-_{j=j^*}(k_x$=$0)=0$. Otherwise, aGR (zCT) is a semiconductor. The twofold band degeneracy of electron states in metallic zCT occurs because there are two zero-energy points, $(0,\xi_{j^*})$ and $(0,2\xi_{j^*})$; equivalently, $(k_x=0,ak_y=2\pi/3)$ and $(k_x=0,ak_y=4\pi/3)$. In Fig.~1, these points are marked by a cross and an open circle. For a semiconducting aGR (zCT), the index $j^*$ of the lowest conduction band can be equal to $(2{\cal N}+1)/3$ or $(2{\cal N}+3)/3$ [$({\cal N}-1)/3$ or  $({\cal N}+1)/3$].  

The next conclusion that follows from the analysis of Eq.~(\ref{c2}) is that near the cross point for aGR and the cross and open-circle points for zCT, the band spectrum can be represented as 
\begin{equation} \label{c4}
\frac{E_{j^*\pm\mu}^{-}}{\Delta_{_{\rm aGR(zCT)}}}= \left \{
\begin{array}{l}
\displaystyle \sqrt{\mu^2\left(1\pm \frac{1}{6}\mu \Delta_{_{\rm aGR(zCT)}}\right)^2 + X^2 },\\
\\
\displaystyle \sqrt{ \left (\pm\mu-\frac{1}{3}\right)^2 +X^2 },  
\end{array} \right.
\end{equation}
where $\Delta_{_{\rm zCT}}=2\Delta_{_{\rm aGR}} =\sqrt{3}\pi/{\cal N} $, $X = \frac{\sqrt{3}}{2}ak_x/\Delta_{_{\rm aGR(zCT)}}$, and $\mu=0,1,...<<\cal N$. The validity of these dispersion relations is ensured by the condition ${\cal N}ak_x<<\mu\ne0$. 

The upper and lower rows in Eq.~(\ref{c4}) respectively refer to the metallic and semiconducting aGR (zCT). Disregarding the term linear in $\mu$ in the upper row yields the previously suggested expression for the band spectrum of $({\cal N}, 0)$ zigzag carbon tubes \cite{Mint} and armchair graphene ribbons \cite{Brey}. However, as can be easily checked by direct calculations of dispersion curves from exact equation (\ref{c2}), this term considerably improves the quality of the approximate description.

{\bf Armchair tubes and zigzag ribbons}.
The spectra of these structures are to be considered separately because the band spectrum of the zigzag GR (zGR) is determined by two equations. An analytic description of the zGR spectrum was repeatedly attempted before but never succeeded.

{\it Armchair tubes}.
For armchair tubes, $ak_x = \kappa_\nu/\sqrt{3}$ plays the role of the transverse quantum number, 
$\kappa_\nu=2\pi \nu/ N$,  $\nu=0,1,...,N-1$, whereas $0\le ak_y\le\pi$ is a continuous variable \cite{Lyuba1}. Equation (\ref{c2}) can then be rewritten as
\begin{equation} \label{c5}
E^\pm_\nu= \sqrt{1+4\cos^2\frac{ak_y}{2}\pm4\left|\cos\frac{ak_y}{2}\cos\frac{\pi\nu}{N}\right|}\,.
\end{equation}
With an exception of the $\nu$ = $0$ band, all other bands are twofold degenerate. Close to the zero energy, that is, for $|q|<<1$, $q=ak_y-2\pi/3$, and $\pi\nu/N<<1$ or $\pi(N$$-$$\nu)/N<<1$, Eq.~(\ref{c5}) is well approximated by
\begin{equation} \label{c6}
E_\nu^{-}(q)= \frac{1}{2}
\sqrt{3q^2+ \left(2\nu\Delta_{_{\rm aCT}}\right)^2\left(1- \frac{\sqrt{3}q}{2}\right)}\,,
\end{equation}
showing that the spectrum  consists of a set of pseudo-parabolic bands with bottoms 
\begin{equation} \label{c7}
E_\nu^{{\rm b}-}= \nu\Delta_{_{\rm aCT}}
 \left[1-\frac{1}{8}\left(\nu\Delta_{_{\rm aCT}}\right)^2\right],
\quad \nu \Delta_{_{\rm aCT}} <<1,
\end{equation}
at $q=q^{\rm aCT}_\nu\equiv\frac{\pi^2\nu^2}{\sqrt{3}N^2}$; here, $\Delta_{_{\rm aCT}} =\pi/N$. The latter quantity would be the band spacing if the term linear in $q$ in the radicand in (\ref{c6}) were disregarded, e.g., as in Ref.~\cite{Mint}. Also worth noting is that for ${\cal N}=\sqrt{3}N$, $\Delta_{_{\rm aCT}}$ = $\Delta_{_{\rm zCT}}$ = $2\Delta_{_{\rm aGR}}$. 

Hence, the spectrum of an aCT contains a nondegenerate band $\nu$ = $0$ with a linear dispersion, followed by a manifold of degenerate $\nu$ and $N-\nu$ bands with a pseudo-parabolic dispersion. This is similar to the case of a metallic zCT. But an important distinction is that in an aCT, there are two propagating states which have different wave vectors, $q<q^{\rm aCT}_\nu$ and $q>q^{\rm aCT}_\nu$ (both values of $k_y$ are positive) but correspond to the same energy. Another difference from a metallic zCT is that the obtained correction to the band spacing, 
$$E_{\nu+1}^{{\rm b}-}-E_\nu^{{\rm b}-} =  \Delta_{_{\rm aCT}} 
\left\{1-\frac{(\Delta_{_{\rm aCT}})^2}{8}[3\nu(\nu+1)+1]\right\},
$$
is not linear but quadratic in $\nu/N$. This correction is small and can therefore be disregarded.

\begin{figure}
\includegraphics[width=0.5\textwidth]{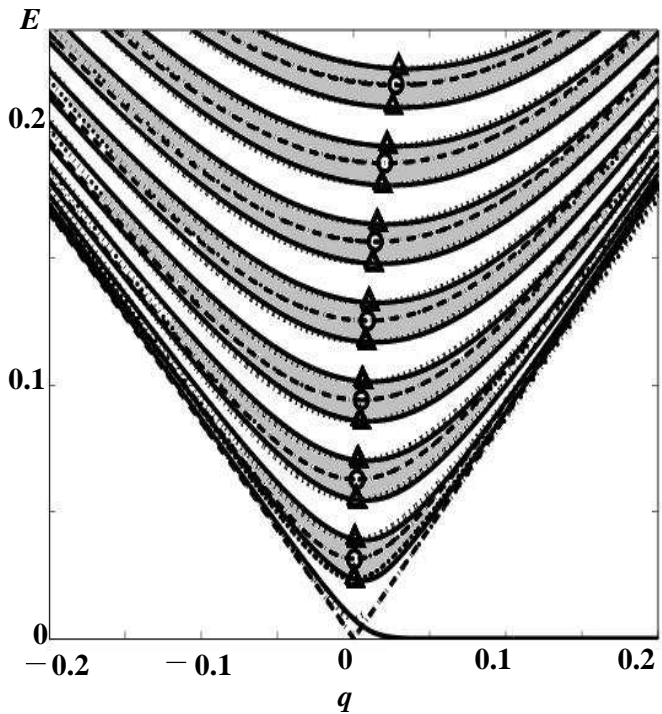}
\caption{One-dimensional bands $\nu =0,1,...,7$ of an $(N,N)$ CT (dashed) and $\nu =0,1,...,14$ of an $N$-wide zigzag GR, $N=100$ (solid lines).  Curves obtained from exact calculations for aCT and zGR are paired with their (often indistinguishable) approximations indicated by dotted lines. The linear dispersion and all curves in the middle of shaded areas demonstrate the aCT band structure (\ref{c6}). The dotted curves correspond to Eqs.~(\ref{c6}), (\ref{c11}) and (\ref{c12}). The band minima calculated in accordance with Eqs.~(\ref{c7}) and (\ref{c13}) are respectively shown by circles and triangles.}
\end{figure}

{\it Zigzag ribbons}.
Cutting an armchair carbon tube along the zigzag direction and "healing" damages of three-coordinated $cp^2$ bonding by hydrogen atoms gives a zigzag graphene ribbon. An essential feature of zigzag ribbons is that discrete values of $0\le ak_x=\kappa_\nu^\pm /\sqrt{3}\le\pi$ are $k_y$-dependent. These values can be found by solving the equation  \cite{Lyuba1}
\begin{equation}\label{c8}
\frac{\sin \kappa^\pm N}{\sin \kappa^\pm(N+1/2)}=\mp2\cos\frac{ak_y}{2},
\end{equation}
where the minus (plus) sign in the right-hand side corresponds to the plus (minus) branch in dispersion relation (\ref{c2}). By combining Eqs.~(\ref{c2}) and (\ref{c8}), $k_y$ can be eliminated. As a result, we obtain
\begin{equation}\label{c9}
E^\pm= \left|\frac{\sin (\kappa^\pm/2)}{\sin \kappa^\pm(N+1/2)}\right|.
\end{equation}
This equation, where only a half of the spectrum with positive energies is presented, substantially simplifies finding the band structure of a zGR. In particular, the extrema of $E^-$, as a function of $\kappa^-$, are given by solutions of the equation
\begin{equation}\label{c10}
\frac{\sin(\kappa^-N)}{\sin \kappa^-(N+1)}= \frac{N}{N+1}.
\end{equation}

For any value of $k_y$, the minus branch of Eq.~(\ref{c8}) has $N$ solutions. One of these solutions is imaginary, $\kappa_0^-=i\delta$, if $k_y$ falls into the interval $2\pi/3+(\sqrt{3}N)^{-1}<ak_y\le\pi$. The energies of such states are within a narrow interval, $E_0^-(q)<(2N+1)^{-1}$, and correspond to states localized near the zigzag edges. In this case, and under the restriction $N\delta>>1$, exact equation (\ref{c9}) is well approximated by 
\begin{equation}\label{c11}
E_0^-(q)= \frac{\sinh \left(-\ln\left[2\sin(\pi/6-q/2)\right]\right)}
{\sinh \left(-\ln\left[2\sin(\pi/6-q/2) \right](2N+1)\right)}.
\end{equation}

The inequality $N\delta>>1$ imposes severe restrictions on the allowed magnitude of $k_y$. However, the above equation describes the dispersion of edge states within a half of the actual interval, $\pi/6<q\le \pi/3$. It can be shown that for small values of $q$, $[\sqrt{3}(N+1/2)]^{-1}\equiv q^c<q<2q^c$, the edge-state dispersion is governed by $E_0^-=\sqrt{3}q\exp\left(-\sqrt{3}Nq\right)$. For the rest of the interval $2\pi/3$ -- $\pi$, i.e., for $2q^c<q\le \pi/6$, exact equations (\ref{c8}) and (\ref{c9}) must be used. Differences between the exact solution of the problem and the approximate description based on the Dirac equation \cite{Brey} are discussed in Ref. \cite{Lyuba2}. 

For $q<q^c$, the dispersion of the lowest energy band in the zGR spectrum is similar to that of the $\nu$ = 0 band in an aCT (see Fig.~2). To obtain an analytic expression for this part and for higher bands, we must use real solutions of Eq.~(\ref{c10}), $\kappa^-_\nu \approx (\nu$+$1/2)\pi/N$, $\nu =0,1,...\,,$ $\nu<<N$. A rather cumbersome procedure of finding these solutions and exploiting them in Eq.~(\ref{c8}) yields 
\begin{equation}\label{c12}
E_\nu^{-}(q)= \frac{1}{2}\sqrt{3q^2 + [(2\nu+1) \Delta_{_{\rm zGR}}]^2\left(1-\frac{\sqrt{3}q}{2}\right)}.
\end{equation}
As explained above, the case $\nu=0$ and $q>0$ is special. 

From Eq.~(\ref{c12}), the bottoms of the $\nu>0$ bands are determined by
\begin{equation} \label{c13}
E_\nu^{{\rm b}-}= \left(\nu+\frac{1}{2}\right)\Delta_{_{\rm zGR}}
\left(1-\frac{1}{8}\left[\left(\nu+\frac{1}{2}\right)\Delta_{_{\rm zGR}}\right]^2\right).
\end{equation}
These energies are attained at 
$q= q^{\rm zGR}_\nu\equiv\frac{\pi^2(\nu+1/2)^2}{4\sqrt{3}N^2}$, $\nu=1,2,...$ . We note that the interband separation in zigzag GRs is well approximated by $ \Delta_{_{\rm zGR}}$ = $\Delta_{_{\rm aCT}}/2$ = $\pi(2N)^{-1}$. Hence, Eq.~(\ref{c13}) can be replaced by $E_\nu^{{\rm b}-}= (\nu+1/2)\Delta_{_{\rm zGR}}$. The ratio between $\Delta_{_{\rm zGR}}$ and $\Delta_{_{\rm aCT}}$ is exactly the same as for zCT--aGR pairs. A comparison of  the approximations in (\ref{c6}), (\ref{c11}), and (\ref{c12}) with the exact calculations (see Fig.~2) proves that these approximations are highly accurate. 

\begin{figure*}
\includegraphics[width=0.95\textwidth]{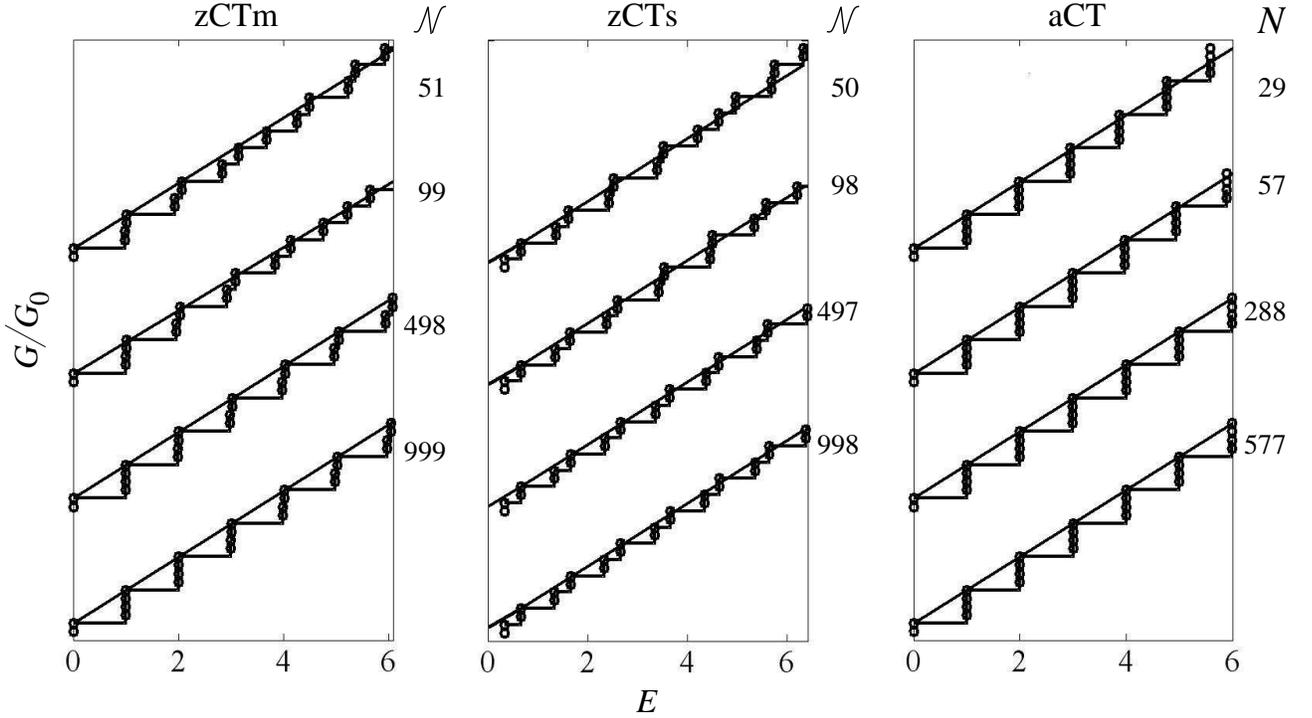}
\caption{Conductance ladders for zigzag metallic (left), semiconducting (middle) and armchair (right) 
carbon tubes with different circumferences; energy scale (from bottom to up): $\frac{\sqrt{3}\pi}{999}\times$ 1, $\frac{999}{498}$,
 $\frac{999}{99}$, $\frac{999}{51}$ on the left panel and similarly on the middle and right panels. 
Markers indicating band opening (band bottom for electrons and band top for holes) are calculated 
from exact dispersion relation (1). For each ladder, the straight line $G(E)/G_0=g_\nu+|E|$ represents an approximate dependence on energy in the limit as ${\cal N}\rightarrow\infty$ (zCT), $N\rightarrow\infty$ (aCT). This provides a visual estimate of the accuracy of relativistic approximation (12), showing that graphene wires do have irregular (left), alternating (middle), and equidistant (right) spectra. Conductance ladders for graphene ribbons have practically the same appearance, except that the step height and width are two times smaller.}
\end{figure*}

By disregarding the term linear in $q$ in Eqs. (\ref{c6}) and (\ref{c12}) and by passing to the new energy scale $\frac{\sqrt{3}}{2}|t|$ ($t$ is the hopping integral), which is more convenient near the Fermi energy, we can express the band structure of graphene ribbons and carbon tubes in a single line 
\begin{equation}\label{c14}
E^\alpha_\nu(k)= 
\sqrt{(m^\alpha_\nu)^2 + k^2 }, 
\end{equation}
where $k$ is the dimensionless wave vector along the ribbon (tube) and the index $\alpha$ specifies the structure and the corresponding expression for $m_\nu^\alpha$. Specifically,
\begin{equation}\label{c15}
 m^\alpha_\nu =
\begin{array}{ll}
\displaystyle \frac{\pi(\nu+1/2)}{\sqrt{3}N},& {\rm zGR},\\ \\
\displaystyle \frac{\pi|\nu|}{{\cal N}}\left(1+ \frac{\pi \nu}{4\sqrt{3}\cal N}\right) ,& {\rm aGRm}, \\\\
\displaystyle \frac{\pi|\nu-1/3|}{{\cal N}} ,& {\rm aGRs},\\\\
\displaystyle \frac{2\pi|\nu|)}{\sqrt{3}N}, & {\rm aCT}, \\\\ 
\displaystyle \frac{2\pi|\nu|}{{\cal N}}\left(1+ \frac{\pi \nu}{2\sqrt{3}\cal N}\right),& {\rm zCTm}, \\\\
\displaystyle \frac{2\pi|\nu-1/3|}{{\cal N}} ,& {\rm zCTs}, 
\end{array}
\end{equation}
where "m" and "s" extensions in labeling indicate metallic and semiconducting aGR (zCT), and $\nu=0,\pm1,...$ for all structures except zGR, where $\nu = 0,1,...$ and the case $\nu=0$, $k>2\pi/3$ is exceptional. 

Equation (\ref{c14}) has the form of a one-dimensional relativistic energy-momentum relation in its conventional representation with the speed of light equal to unity. This is not more than a formal analogy.

As follows from Eqs.~(\ref{c14})  and (\ref{c15}), the spectra of achiral graphene ribbons and carbon tubes can be classified into three groups: (i) metallic spectra with equally spaced bands, as for the aCT and zGR; (ii) metallic spectra with a regularly irregular band spacing, as for the zCT and aGR; and (iii) semiconducting spectra with the band spacing alternating between $\frac{\pi}{3\cal N}$ and $\frac{2\pi}{3\cal N}$ (between $\frac{2\pi}{3\cal N}$ and $\frac{4\pi}{3\cal N}$), as for semiconducting armchair ribbons (zigzag tubes). 

\section{Conductance} 
In the framework of the Landauer approach \cite{Land,But}, the zero-bias, zero-temperature conductance of an ideal wire is equal to
\begin{equation}\label{c16}
G(E)= G_0\sum_{\nu}g_\nu T_{\nu}(E),
\end{equation}
where $G_0=2e^2/h$ is the conductance quantum, $g_\nu$ is the degeneracy of band states (spin degeneracy 2 is included into $G_0$), and the transmission coefficient $T_\nu$ is zero or unity, depending on whether the $\nu$th band is open for charge carriers or not. 

For the band structure specified in Eq. (\ref{c14}), $T_\nu(E) = \Theta(E-m^\alpha_\nu)$ for conduction bands and $T_\nu(E) = \Theta(|E-m^\alpha_\nu|)$ for valence bands, where $\Theta(x)$ is the Heaviside step function. Thus, the quantum conductance as a function of energy has the form of a ladder, symmetrically ascending with an increase in $E$ for electrons, and with a decrease in the energy for holes. The height of the $\nu$th ladder step is determined by the band-state degeneracy $g_{\nu\rm=0}=$ 1 (2) for zGR (aCT), otherwise, $g_\nu=2$ (4) for GR (CT). Three types of the band spectrum identified above can be translated into three corresponding types of conductance ladders with regular (aCT and zGR), irregular (zCTm and aGRm), and alternating (zCTs and aGRs) width of steps. 

Needless to say, the appearance of $G(E)$ illustrated in Fig.~3 depends on the energy scale determined by the ribbon width (tube circumference). It can vary from the ladder-shaped curve, ascending with the increase of $|E|$, to a straight line $G=G_0(g_\nu+|E|)$. The latter dependence shows that metallic and semiconducting graphene-based wires are indistinguishable as classic conductors.
Furthermore, if the irregularity of the step width is resolved, the difference between conductance ladders
 of metallic zCT (aGR) and aCT (zGR) is not questioned, but they look identically in a larger energy scale. This can be seen in Fig.~3, which illustrates conductance in different energy scales. 

The above consideration concerns four basic graphene wires which are discussed in more detail elsewhere \cite{O}.
It provides a useful reference for calculations of electron and hole coherent transmission in various types of graphene contacts.\\\\
This work was supported by the Special Program of the Section of 
Physics and Astronomy of NANU, Visby program (SI), and an individual SI grant for A.O.


\begin{thebibliography}{99}

\bibitem{Geim}
A.K. Geim and K.S. Novoselov, {\it Nature materials} {\bf 6}, 183 (2007).

\bibitem{Peres1}
N.~M.~R. Peres, A.~H. Castro Neto, and F. Guinea,
 Phys. Rev. B {\bf 73}, 195411 (2006). 

\bibitem{Lyuba1}
L. Malysheva and A. Onipko, Phys. Rev. Lett. \textbf{100}, 186806 (2008).

\bibitem{Fujita}
M. Fujita, K. Wakabayashi, K. Nakada, and K. Kusakabe, J. Phys. Soc. Japan \textbf{65}, 1920 (1996). 

\bibitem{Mint}
J.W. Mintmire and C.T. White, Phys. Rev. Lett. \textbf{B 81}, 2506 (1998).

\bibitem{Brey}
L. Brey and H.\,A. Fertig, Phys. Rev. \textbf{B 73}, 235411 (2006).

\bibitem{Ch}
H. Zheng, Z.\,F. Wang, T. Luo, Q.\,W. Shi, and J. Chen, Phys. Rev. \textbf{B 75}, 165414 (2007).

\bibitem{Lyuba2}
L. Malysheva and A. Onipko, arXiv:0802.1385v2 [cond-mat.mes-hall].

\bibitem{Wallace}
P.~R. Wallace,  Phys. Rev. {\bf 71}, 622 (1947).

\bibitem{Saito}
R.Saito, G. Dresselhaus, and M.S. Dresselhaus, {\it Physical Properties of Carbon Tubes}, Imperial College, London (1998).

\bibitem{Land}
R. Landauer, IBM J. Res. Dev. \textbf{1}, 233 (1957); \textbf{32}, 306 (1988).

\bibitem{But}
M. B\"uttiker, Phys. Rev. Lett. \textbf{57}, 1761 (1986).

\bibitem{O}
A. Onipko, Phys. Rev. B \textbf{78}, 245412 (2008).


\end{thebibliography}
\end{document}